\documentclass[12pt]{article}
\usepackage[latin1]{inputenc}
\usepackage{slashed}
\usepackage{amsmath}
\usepackage{textcomp}
\usepackage{amssymb}
\usepackage{amsfonts}
\usepackage{indentfirst}
\usepackage{color}
\usepackage[top=2cm,bottom=2cm,left=3cm,right=2cm]{geometry}
\newcommand{\nin}{\noindent}
\newcommand{\be}{\begin{equation}}
\newcommand{\ee}{\end{equation}}
\newcommand{\bea}{\begin{eqnarray}}
\newcommand{\eea}{\end{eqnarray}}

\newcommand{\nn}{\nonumber\\}

\newcommand{\ol}{\overline}

\begin{document}

\begin{flushleft} 
KCL-PH-TH/2014-{\bf23} 
\end{flushleft}

\vspace{1cm}

\begin{center}
 
{\Large{\bf Non-Hermitian Lagrangian for \\ quasi-relativistic fermions}}
 
\vspace{0.5cm}

Jean Alexandre

{\small King's College London, Department of Physics, WC2R 2LS, UK}

\vspace{2cm}

{\bf Abstract}

\end{center}

\nin We present a Lorentz-symmetry violating Lagrangian for free fermions, which is local but not Hermitian, 
whereas the corresponding 
Hamiltonian is Hermitian but not local. A specific feature of the model is that the dispersion relation is relativistic 
in both the IR and in the UV,
but not in an intermediate regime, set by a given mass scale. The consistency of the model is shown by the study of 
properties expected in analogy with the Dirac Lagrangian.

\section{Introduction}

Lorentz-invariance violation (LIV) has had an increasing attention these last years, and provides opportunities to develop quantum gravity phenomenology.
A generic bottom-up approach consists in the SME (Standard Model Extension) \cite{SME} (see \cite{SMEreview} for a review), 
where tensors of different rank are allowed to acquire non-trivial vacuum
expectation values, therefore breaking either invariance under 3-dimensional rotation and/or Lorentz boosts. 
A possible top-down approach consists in deriving LIV operators from gravitational models, and an example 
is given by a brane model \cite{nem} 
where LIV effects arise, in the effective theory, from the choice of a specific frame, where bulk topological defects are slow-moving. 
Such a model can give a microscopic origin 
of LIV Lagrangians as the ones studied in \cite{mdyn1} and \cite{mdyn2}, where the dynamical generation of masses and flavour oscillations result from LIV 
kinematics.

We propose here a new LIV Lagrangian for free fermions, which is not Hermitian but leads to a Hermitian Hamiltonian, 
such that the spectrum is real. On the other hand, the Hamiltonian is not local, whereas the Lagrangian is. The reason is that
the non-Hermitian term in the Lagrangian contains mixed derivatives $\partial_k\partial_0\psi$, and since the Hamiltonian operator $H$ is
defined by $i\partial_0\psi=H\psi$, it involves the inverse of a quantity which contains space derivatives. The Hamiltonian can therefore be
seen as a resummation of an infinite series in space derivatives, and is thus not local.
Therefore the possibility to have a Hermitian but non-local Hamiltonian is a consequence of Lorentz-symmetry violation, where
mixed derivatives can occur in the Lagrangian.
Note that consistent models with non-Hermitian Hamiltonians are known, and give rise to a whole area of study, based on parity and time
reversal (PT) symmetry \cite{bender}. 

The specific form 
of this Lagrangian allows quasi-relativistic kinematics in the sens that, in both the infrared (IR) and ultraviolet (UV) regimes, 
the
dispersion relation is relativistic. It differs from the usual relativistic form only in an intermediate regime, characterised by 
the LIV scale $M$.
For this reason, the present model does not improve the convergence of Feynman graphs, unlike Lifshitz theories (see \cite{Lifreview} 
for a review in 
Particle Physics). Also,
because the Hamiltonian is not local, the energies can be though of the resummation of an infinite series 
in the momentum, although the Lagrangian contains a finite number of space derivatives.

Sections 2 and 3 study kinematic and dynamical aspects of the model respectively, and section 4.
suggests an extension to simple dynamics, involving a Yukawa interaction, which shows that fermion dynamical 
mass generation can be studied perturbatively in this model.

\section{Kinematics}

We introduce here a model which can be derived from the fermionic sector of the SME \cite{fermsec}, 
with a specific set of tensor vacuum expectation values.
We describe fundamental features which are necessary to check the consistency: {\it(i)} the equation of motion, for which 
the derivation is not trivial in the situation of a non-Hermitian Lagrangian; {\it(ii)} the conserved current, which is not 
the same as in the case of the Dirac equation, and which is necessary to have a unitary theory. Also, we show that the concept 
of helicity is ambiguous, since the usual right- and left-handed components are always coupled, even in the massless case.

In what follows we use the metric $\eta_{\mu\nu}=$ diag(1,-1,-1,-1), such that the Laplacian operator is $\Delta=-\partial_k\partial^k$, 
and we note
$\vec\gamma\cdot\vec\partial=-\gamma^k\partial_k$.

\subsection{Lagrangian}

We consider the Lorentz-symmetry violating free Lagrangian
\be\label{model}
\mathcal{L} = \ol\psi \left(1-i\frac{\vec\gamma\cdot\vec\partial}{M}\right)i\slashed{\partial} \psi -m\ol\psi\psi ~,
\ee
where $M$ is the LIV mass scale. This Lagrangian is not Hermitian, since it can be written 
\be
\mathcal{L} = \ol\psi\left(i\slashed\partial+\frac{\Delta}{M}\right)\psi+\frac{1}{M}\ol\psi~\vec\gamma\cdot\vec\partial~
\gamma^0\partial_0~\psi-m\ol\psi\psi ~,
\ee
and contains the anti-Hermitian operator 
\be\label{anti}
\ol\psi~\vec\gamma\cdot\vec\partial~\gamma^0\partial_0~\psi
=-(\ol\psi~\vec\gamma\cdot\vec\partial~\gamma^0\partial_0~\psi)^\dagger
~+~\mbox{total derivatives}~.\nonumber
\ee
An important point in the Lagrangian (\ref{model}) is the absence of higher order time derivatives. This prevents the appearance of new poles in energies 
in the propagator, which can be seen in subsection 4.1. Therefore this model does not introduce ghost particles, 
which is a feature common with Lifshitz-type models.

\subsection{Equation of motion}

We show now that the equation of motion is identical to the one obtained from the usual procedure, which consists in varying the action with respect to
$\ol\psi$, keeping $\psi$ constant. This is not trivial for a non-Hermitian Lagrangian, since in this case the variation of the action with respect
to $\psi$ does not lead to the previous equation of motion after taking the Hermitian conjugate.

We consider here the Majorana representation for gamma matrices, where these are all imaginary. The four fermion components
are written $\psi_a=\phi_a+i\chi_a$, where $a=1,2,3,4$ and $(\phi_a,\chi_a)$ are real. The action can be written
\bea
S&=&\int \ol\psi\left(i\slashed\partial-m+\frac{1}{M}\vec\gamma\cdot\vec\partial~\slashed\partial\right)\psi\\
&=&\int (\phi_b-i\chi_b)\left(i\gamma^0\slashed\partial-m\gamma^0
+\frac{1}{M}\gamma^0\vec\gamma\cdot\vec\partial~\slashed\partial\right)_{bc}(\phi_c+i\chi_c)~,
\eea
and the equations of motion are obtained by setting the variations of $S$ with respect to $\phi_a$ and $\chi_a$ to 0.
We obtain 
\bea
\frac{\delta S}{\delta\phi_a}
&=&\left(i\gamma^0\slashed\partial-m\gamma^0+\frac{1}{M}\gamma^0\vec\gamma\cdot\vec\partial~\slashed\partial\right)_{ac}(\phi+i\chi)_c\\
&&-\left(-i\gamma^0\slashed\partial-m\gamma^0+\frac{1}{M}\gamma^0\vec\gamma\cdot\vec\partial~\slashed\partial\right)_{ba}(\phi-i\chi)_b\nn
&=&\left(i\gamma^0\slashed\partial+i\left(\gamma^0\slashed\partial\right)^T-m\gamma^0+m\gamma^{0T}
+\frac{1}{M}\gamma^0\vec\gamma\cdot\vec\partial~\slashed\partial
-\frac{1}{M}(\gamma^0\vec\gamma\cdot\vec\partial~\slashed\partial)^T\right)_{ac}\phi_c\nn
&&+\left(-\gamma^0\slashed\partial+\left(\gamma^0\slashed\partial\right)^T-im\gamma^0-im\gamma^{0T}
+\frac{i}{M}\gamma^0\vec\gamma\cdot\vec\partial~\slashed\partial
+\frac{i}{M}(\gamma^0\vec\gamma\cdot\vec\partial~\slashed\partial)^T\right)_{ac}\chi_c~
\nonumber.
\eea
In the Majorana representation, we have
\be
\gamma^{\mu T}=-\gamma^{\mu\dagger}=-\gamma^0\gamma^\mu\gamma^0~,
\ee
such that
\be
\frac{\delta S}{\delta\phi_a}
=2\left(\gamma^0(i\slashed\partial-m)\phi+\frac{i}{M}\gamma^0\vec\gamma\cdot\vec\partial~\slashed\partial\chi\right)_a
+\frac{2}{M}(\gamma^0\Delta\psi^\star)_a~,
\ee
Similar steps lead to
\be
\frac{\delta S}{\delta\chi_a}
=2\left(\gamma^0(i\slashed\partial-m)\chi-\frac{i}{M}\gamma^0\vec\gamma\cdot\vec\partial~\slashed\partial\phi\right)_a
+\frac{2i}{M}(\gamma^0\Delta\psi^\star)_a~,
\ee
and it is easy to see that the term involving the Laplacian $\Delta\psi^\star$ cancels in the following linear combination
\be
\frac{1}{2}\frac{\delta S}{\delta\phi_a}+\frac{i}{2}\frac{\delta S}{\delta\chi_a}=
\left(\gamma^0\left(i\slashed\partial-m+\frac{1}{M}\gamma^0\vec\gamma\cdot\vec\partial~\slashed\partial\right)(\phi+i\chi)\right)_a~.
\ee
But the latter equation can also be written
\be
\gamma^0\frac{\delta S}{\delta\psi^\star}=\left(i\slashed\partial-m+\frac{1}{M}\gamma^0\vec\gamma\cdot\vec\partial~\slashed\partial\right)\psi~,
\ee
which corresponds to the variation $\delta S/\delta\ol\psi$ performed when $\ol\psi$ and $\psi$ are considered independent, such that 
the equation of motion is finally 
\be\label{equamot}
\left(1-i\frac{\vec\gamma\cdot\vec\partial}{M}\right)i\slashed\partial\psi=m\psi~,
\ee
and is not given by the Hermitian conjugate of $\delta S/\delta\psi=0$

\subsection{Helicity and conserved current}

The concept of helicity is ambiguous for the Lagrangian (\ref{model}), since the kinetic term mixes both helicities. 
Indeed, it is easy to see that
\bea
&&\ol\psi~\vec\gamma\cdot\vec\partial~\slashed\partial\psi=\ol\psi_R\vec\gamma\cdot\vec\partial~\slashed\partial\psi_L
+\ol\psi_L\vec\gamma\cdot\vec\partial~\slashed\partial\psi_R\\
\mbox{where}~&&~\psi_L=\frac{1}{2}(1-\gamma^5)\psi~~,~~\psi_R=\frac{1}{2}(1+\gamma^5)\psi~,\nonumber
\eea
such that helicity is not conserved, even in the massless case.

The conserved current is obtained in the usual way. The equation of motion (\ref{equamot}) leads to
\be\label{formal}
i\slashed\partial\psi=m\left(1-i\frac{\vec\gamma\cdot\vec\partial}{M}\right)^{-1}\psi
=m~\frac{1+i\vec\gamma\cdot\vec\partial/M}{1-\Delta/M^2}~\psi~,
\ee
and multiplying by $\ol\psi$ on the left, one obtains
\be\label{step1}
i\ol\psi\slashed\partial\psi=m\ol\psi~\frac{1+i\vec\gamma\cdot\vec\partial/M}{1-\Delta/M^2}~\psi~.
\ee
Then, one takes the Hermitian conjugate of the equation (\ref{formal}), and multiplies it by $\gamma^0\psi$ on the right, 
to obtain
\be\label{step2}
-i\ol\psi\overleftarrow{\slashed\partial}\psi=m\ol\psi~\frac{1-i\vec\gamma\cdot\overleftarrow\partial/M}
{1-\overleftarrow\Delta/M^2}~\psi~.
\ee
The difference of eqs.(\ref{step1}) and (\ref{step2}) gives
\be\label{difference}
i\partial_\mu(\ol\psi\gamma^\mu\psi)=m\ol\psi\left(\frac{1+i\vec\gamma\cdot\vec\partial/M}{1-\Delta/M^2}
-\frac{1-i\vec\gamma\cdot\overleftarrow\partial/M}{1-\overleftarrow\Delta/M^2}\right)\psi~,
\ee
where the right-hand side is actually a total derivative. To see this, we first define
\be
\xi\equiv\frac{1}{1-\Delta/M^2}\psi=\sum_{n=0}^\infty\left(\frac{\Delta}{M^2}\right)^n\psi~,
\ee
and the right-hand side of the equation (\ref{difference}) reads then
\bea
&&\frac{m}{M}\ol\xi\left(i(\vec\gamma\cdot\overleftarrow\partial+\vec\gamma\cdot\vec\partial)
+\frac{1}{M}(\Delta-\overleftarrow\Delta)
-\frac{i}{M^2}(\vec\gamma\cdot\overleftarrow\partial~\Delta+\vec\gamma\cdot\vec\partial~\overleftarrow\Delta)\right)\xi\\
&=&-\frac{m}{M}\partial_l\left(i\ol\xi\gamma^l\xi+\frac{1}{M}(\ol\xi\partial^l\xi-\partial^l~\ol\xi~\xi)
-\frac{i}{M^2}(\ol\xi(\gamma^k\partial_k)\partial^l\xi+\partial^l(\partial_k\ol\xi)\gamma^k\xi
-\partial_k\ol\xi~\gamma^l~\partial^k\xi)\right)\nonumber~.
\eea
The conserved current is therefore of the form ($l=1,2,3)$
\bea
j^\mu&=&\ol\psi\gamma^\mu\psi
+\frac{m}{M}\eta^{\mu l}\left(\ol\psi\gamma_l\psi-\frac{i}{M}(\ol\psi\partial_l\psi-\partial_l~\ol\psi~\psi)+\cdots\right)\nn
&=&\ol\psi\gamma^\mu\psi+\frac{m}{M}\eta^{\mu l}~\ol\psi~\overleftrightarrow{\cal O}_l~\psi~,
\eea
where ${\cal O}_l$ is an operator containing space derivatives $\partial_l$ and the gamma matrices $\gamma_l$.
Similarly to the Hamiltonian operator (\ref{hamilton}), the conserved current 
is not local, but the probability density is the same as for the Dirac equation: $j^0=\ol\psi\gamma^0\psi=\psi^\dagger\psi$.\\

\section{Energetics}

We show here that the Hamiltonian associated to the model (\ref{model}) is Hermitian, and it therefore leads to real energies.
The main point is to exhibit a new type of dispersion relation, which is relativistic in both the IR and the UV, 
these two regimes being separated by $M$. The intermediate regime, for energies of the order $M$, shows a
departure from a relativistic dispersion relation, and this specific feature will allow a non-trivial dynamical mass generation,
as explained in the next section. The Fouldy-Wouthuysen transformation
is obtained in a similar way as in the Dirac case, and it has the advantage to
exhibits the negative- and positive-energy modes of the plane wave solutions.

\subsection{Hamiltonian}

Although the Lagrangian (\ref{model}) is not Hermitian, we show here that the corresponding Hamiltonian is Hermitian.
The Schrodinger form $i\partial_0\psi=H\psi$ of the equation of motion (\ref{formal}) leads to the identification of the 
Hamiltonian operator $H$
\be\label{hamilton}
H=\frac{m\gamma^0}{1-\Delta/M^2}+\gamma^0\left(1+\frac{m/M}{1-\Delta/M^2}\right)i\vec\gamma\cdot\vec\partial~,
\ee
which gives the Hermitian Hamiltonian density $\psi^\dagger H\psi$.
As a consequence the spectrum must be real, as will be seen in the next paragraph.
We note that the Hamiltonian (\ref{hamilton}) is not local, since it can be understood as the resummation 
of an infinite series in $\Delta$, whereas the Lagrangian 
(\ref{model}) contains a finite number of derivatives. 
As explained in the introduction, this is a consequence of the mixed derivative term $\vec\gamma\cdot\vec\partial\gamma^0\partial_0\psi$ 
in the Lagrangian, which is also the reason for which the usual definition
\be
{\cal H}=\frac{{\cal L}\overleftarrow\partial}{\partial(\partial_0\psi)}\partial_0\psi-{\cal L}~,
\ee
cannot be used to determine the Hamiltonian density, since 
\be
{\cal H}=\ol\psi\left(i\vec\gamma\cdot\vec\partial-\frac{\Delta}{M}+m\right)\psi~\ne~\psi^\dagger H\psi~.
\ee

\subsection{Dispersion relation}

The dispersion relation for the Lagrangian (\ref{model}) is obtained by plugging a plane wave into the equation of motion (\ref{equamot}), 
which leads to
\be\label{equamotFourier}
\left(1-\frac{\vec p\cdot\vec\gamma}{M}\right)(\omega\gamma^0-\vec p\cdot\vec\gamma)\psi=m~\psi~,
\ee
and hence
\be
(\omega\gamma^0-\vec p\cdot\vec\gamma)\psi=m~\frac{1+\vec p\cdot\vec\gamma/M}{1+p^2/M^2}~\psi~.
\ee
We therefore have
\be
\left[\omega\gamma^0-\vec p\cdot\vec\gamma\left(1+\frac{m/M}{1+p^2/M^2}\right)-\frac{m}{1+p^2/M^2}\right]\psi=0~,
\ee
such that
\be\label{disprel}
\omega^2=\frac{m^2}{(1+p^2/M^2)^2}+p^2\left(1+\frac{m/M}{1+p^2/M^2}\right)^2~.
\ee
We note that $\mathcal{L}^\dagger$ leads to the same dispersion relation as $\mathcal{L}$.
Also, this dispersion relation is relativistic for $m=0$, and for $m\ne0$,
it is ``quasi-relativistic'' in the sense that, in both IR and UV regimes, it has a relativistic form 
\bea
\omega^2 &\simeq& m^2+p^2~~~~~\mbox{for}~~p<<M\\
\omega^2 &\simeq& p^2~~~~~\mbox{for}~~p>>M\nonumber~.
\eea
The dispersion relation (\ref{disprel}) deviates from relativistic kinematics in the intermediate regime $p\sim M$ only,
which is an important difference with Lifshitz-type models,
for which the UV regime is characterised by $\omega^2\sim p^{2+n}$ with $n>0$. \\
The product of phase and group velocities is 
\be\label{bound}
v_pv_g=\frac{\omega}{p}\frac{d\omega}{dp}=1+\frac{m}{M}~ \frac{2-m/M}{(1+p^2/M^2)^2}~,
\ee
and shows that fermions described by the Lagrangian (\ref{model}) are superluminal if one assumes that $m<2M$.
Nevertheless, for a typical Standard Model mass $m$ and a typical Grand Unified Theory mass $M$, the Lorentz-symmetry violating upper bound, 
which is of the order $|v_pv_g-1|\lesssim10^{-15}$ for electrons \cite{upper}, is satisfied for any momentum $p$.\\
Finally, we note that for $m=2M$, we have the exact relativistic relation $v_pv_g=1$, although
the dispersion relation (\ref{disprel}) is not relativistic.

\subsection{Foldy-Wouthuysen transformation}

This field transformation is helpful for non-relativistic approximations \cite{FW}, and consists in writing the 
equation of motion in the form
\be\label{equamotchi}
i\partial_0\chi(t,p)=\omega(p)\gamma^0\chi(t,p)~,
\ee 
where $\omega(p)$ is the energy obtained from the dispersion relation (\ref{disprel}), and $\chi=U\psi$ with $U$ a unitary
matrix. As usual for this transformation, one looks for $U$ in the form
\be
U\equiv\exp\left(\theta\frac{\vec p\cdot\vec\gamma}{p}\right)=\cos\theta+\frac{\vec p\cdot\vec\gamma}{p}\sin\theta~,
\ee
where $p=\sqrt{p^2}$, and $U^{-1}$ is obtained by changing $\theta$ into $-\theta$. The latter angle is to be determined,
in order to obtain the equation of motion (\ref{equamotchi}), which is the aim of this paragraph.\\
The Hamiltonian (\ref{hamilton}) is, in Fourier components,
\be
H=\frac{m\gamma^0}{1+p^2/M^2}+\gamma^0\vec p\cdot\vec\gamma\left(1+\frac{m/M}{1+p^2/M^2}\right)~,
\ee
and a straightforward calculation leads to
\bea
UHU^{-1}&=&\gamma^0\left[\vec p\cdot\vec\gamma\left(A\cos(2\theta)-\frac{B}{p}\sin(2\theta)\right)+B\cos(2\theta)
+Ap\sin(2\theta)\right]\nn
\mbox{where}~&&A=1+\frac{m/M}{1+p^2/M^2}~~~\mbox{and}~~~B=\frac{m}{1+p^2/M^2}~.\nonumber
\eea
The vanishing of the coefficient of $\vec p\cdot\vec\gamma$ imposes 
\be\label{theta}
\tan(2\theta)=\frac{Ap}{B}=\frac{p}{m}\left(1+\frac{m}{M}+\frac{p^2}{M^2}\right)~,
\ee
and it is easy to see that one is left with the expected form
\be
UHU^{-1}=B\cos(2\theta)+Ap\sin(2\theta)=\omega\gamma^0~,
\ee
where $\omega$ is obtained from the dispersion relation (\ref{disprel}). As expected, the angle defined by eq.(\ref{theta})
leads to the usual Foldy-Wouthuysen result $\tan(2\theta)_{M\to\infty}=p/m$.\\
We note that the FW transformation for a Standard Model Extension Lagrangian is studied in \cite{SMEFW}, where the Authors consider all
the possible CPT and Lorentz-symmetry violating terms as an extension of the Dirac Lagrangian. \\

\section{Perturbative dynamical mass}

We present here a physical consequence of the model (\ref{model}), which is 
the perturbative generation of a fermion mass, when the bare mass vanishes $m=0$. 
Generating a fermion mass perturbatively is usually not possible, and one needs to use non-perturbative approaches 
to generate such a mass dynamically. The Schwinger-Dyson approach is an example, and consists in summing an infinite set of Feynman graphs,
to lead to a dynamical mass which is not analytic in the coupling constant.

\subsection{Propagator}

The propagator $S$ for the model (\ref{model}) is defined by
\be
\left[\left(1-\frac{\vec p\cdot\vec\gamma}{M}\right)(\omega\gamma^0-\vec p\cdot\vec\gamma)-m\right](-iS)=1~,
\ee
hence
\be
\left[\omega\gamma^0-\vec p\cdot\vec\gamma\left(1+\frac{m/M}{1+p^2/M^2}\right)-\frac{m/M}{1+p^2/M^2}\right](-iS)
=\frac{1+\vec p\cdot\vec\gamma/M}{1+p^2/M^2}~,
\ee
such that finally
\be\label{prop}
S=i~\frac{\omega \gamma^0 -\vec{p}\cdot\vec{\gamma} \left(1+\frac{m/M}{1+p^2/M^2}\right)+\frac{m}{1+p^2/M^2}}
{\omega^2-p^2\left(1+\frac{m/M}{1+p^2/M^2}\right)^2-\frac{m^2}{(1+p^2/M^2)^2}}~\frac{1+ \vec p \cdot \vec\gamma/M}{ 1+p^2/M^2}~.
\ee 
It is interesting to note that for massless fermions $m=0$, the propagator (\ref{prop}) has a non-vanishing trace
\be
\frac{1}{4}\mbox{tr}\left\{S_{m=0}\right\}=\frac{i~p^2/M}{(1+p^2/M^2)(\omega^2-p^2)}~,
\ee
which allows the perturbative generation of a fermion mass, as described bellow.\\

\subsection{Yukawa interaction}

We introduce here a toy-model Yukawa interaction with a real and Lorentz-invariant scalar field $\phi$:
\be
{\cal L}'=\ol\psi \left(1-i\frac{\vec\gamma\cdot\vec\partial}{M}\right)i\slashed{\partial} \psi
+\frac{1}{2}\partial_\mu\phi\partial^\mu\phi-\frac{m^2_H}{2}\phi^2-g\phi\ol\psi\psi~,
\ee
for which the one-loop fermion mass is given by
\bea
m_f^{(1)}&=&i\frac{4\pi g^2}{(2\pi)^4}\int_{-\infty}^\infty d\omega\int_0^\Lambda p^2dp
\frac{p^2/M}{(1+p^2/M^2)(\omega^2-p^2)(\omega^2-p^2-m_H^2)}\\
&=&\frac{g^2M}{4\pi^3\mu^2}\int_0^{\Lambda/M}\frac{x^4~dx}{1+x^2}\left(\frac{1}{x}-\frac{1}{\sqrt{x^2+\mu^2}}\right)~,\nonumber
\eea
where $\Lambda$ is a cut off and $\mu\equiv m_H/M$. An expansion for $\mu<<1$ gives the result
\be
m_f^{(1)}\simeq \frac{g^2M}{16\pi^3}\ln\left(1+\frac{\Lambda}{M}\right)~.
\ee
If we assume a Planckian mass for the cut off $\Lambda$ and a Grand Unified Theory mass for $M$, then we can see that a Yukawa 
coupling which satisfies $g^2\sim M/\Lambda$ can be consistent with a neutrino mass $m^{(1)}$. Such a small Yukawa coupling
could be natural in the context of a quantum-gravity-induced neutrino mass, as done in \cite{pilaftsis} for example.
A coupling which scales as the inverse of the cut off has been used in \cite{mdyn2} to describe the 
Lorentz-symmetric limit of LIV models which allow the dynamical generation of flavour oscillations.

\section{Conclusion}

This article shows that a non-Hermitian Lagrangian is mathematically consistent and could lead to new physics. 
The relevance of non-Hermitian Lagrangians to Physics is a quite recent area of research \cite{bender}, and is currently 
being developed in different areas of quantum field theory \cite{PTQFT}.
We note the work \cite{BJR}, which
also involves a consistent non-Hermitian fermionic Lagrangian, featuring a parity-violating mass term, which is invariant under simultaneous
parity and time reversal though.
The latter model is studied also through the Foldy-Wouthuysen transformation in \cite{AB}. It is expected that further 
studies in this direction will generate a whole new are of Physics, which could be relevant to beyond the Standard Model.

If one wishes to gauge the present model though, one cannot respect both gauge invariance and renormalizability, 
unless in the $z=2$ Lifshitz context, where a vortex attached to two gauge and two fermion propagators is allowed. But the Yukawa interaction
considered in the present article could be relevant to neutrinos-Higgs interactions. 
The coupling of neutrinos to weak bosons would then not need to respect gauge invariance, at least after the electro-weak spontaneous 
symmetry breaking, and the corresponding extension of the Standard Model would then be possible in the context of effective theories.

\vspace{1cm}

\nin{\bf Acknowledgements} I would like to thank Carl Bender for illuminating discussions, and Julio Leite for relevant comments.

\end{document}